\documentclass[prl,amsmath,amssymb,dvips,twocolumn,showpacs]{revtex4}
\usepackage{epsfig}
\usepackage{bm}

\newcommand{\beq}{\begin{equation}}
\newcommand{\eeq}{\end{equation}}
\newcommand{\bea}{\begin{eqnarray}}
\newcommand{\eea}{\end{eqnarray}}
\newcommand{\Tr}{\text{Tr}}

\begin{document}

\title{Momentum Distribution and Contact of the Unitary Fermi gas}

\author{Joaqu\'{\i}n E. Drut$^{1,2}$, Timo A. L\"ahde$^3$, and Timour Ten$^{1,4}$ }
\affiliation{$^1$Theoretical Division, Los Alamos National Laboratory, Los Alamos, New Mexico 87545--0001, USA}
\affiliation{$^2$Department of Physics, The Ohio State University, Columbus, Ohio 43210--1117, USA}
\affiliation{$^3$Helsinki Institute of Physics and Department of Applied Physics, 
Aalto University, FI-00076 Aalto, Espoo, Finland}
\affiliation{$^4$Department of Physics, University of Illinois, Chicago, Illinois 60607--7059, USA}

\date {\today}

\begin{abstract}
We calculate the momentum distribution $n(k)$ of the Unitary Fermi Gas using Quantum Monte Carlo 
calculations at finite temperature $T/\epsilon_F^{}$ as well as in the ground state. At large momenta $k/k_F^{}$, we find that $n(k)$
falls off as $C/k^{4}_{}$, in agreement with the Tan relations. From the asymptotics of $n(k)$, 
we determine the contact~$C$ as a function of $T/\epsilon_F^{}$ and present a 
comparison with theory. At low~$T/\epsilon_F^{}$, we find that $C$ increases with temperature, and we tentatively
identify a maximum around $T/\epsilon_F^{} \simeq 0.4$. Our calculations are performed on lattices of spatial 
extent up to $N_x^{} = 14$ with a particle number per unit volume of $\simeq 0.03 - 0.07$.
%, which corresponds to $\simeq 45$ particles for $N_x^{} = 10$ and $\simeq 75$ particles for $N_x^{} = 12$. 
\end{abstract}


\maketitle

%%%%%%%%%%%%%%%%%%%

The Unitary Fermi Gas~(UFG) is one of the most interesting
strongly interacting systems known to date, as it saturates the unitarity bound on the
quantum mechanical scattering cross section $\sigma_0^{} \leq 4\pi / k^2$. Since the proposal of the UFG as a model for dilute neutron
matter by Bertsch~\cite{GFBertsch} and its realization in ultracold atom experiments~\cite{FirstExperiments}, the UFG has garnered 
widespread attention across multiple disciplines, including atomic physics~\cite{Atoms}, nuclear structure~\cite{Nuclei} 
and relativistic heavy-ion collisions~\cite{RHIC}. The UFG is defined as a two-component many-fermion system
in the limit of short interaction range $r_0^{}$ and large $s$-wave scattering length~$a$,
\beq
0 \leftarrow k^{}_F r^{}_0 \ll 1 \ll k^{}_F a \rightarrow \infty,
\eeq
with $k^{}_F \equiv (3 \pi^2 n)^{1/3}_{}$ the Fermi momentum and $n$ the particle number density. The special properties
of the UFG arise from the fact that it is characterized by a single scale, given by the inter-particle distance $\sim k_F^{-1}$,
without reference to the details of the interaction.
While the thermodynamic properties of the UFG are {\it universal}~\cite{Universality}, the lack of an 
obvious dimensionless expansion parameter makes the UFG a challenging many-body problem. 

In spite of the challenges of the unitary limit, much progress has been made with purely analytical methods. Notably, in 2005 Tan 
was able to derive exact thermodynamic relations~\cite{ShinaTan} in terms of a universal quantity known as the ``contact" $C$, which
determines the number of pairs separated by short distances. Since then, the Tan relations have been re-derived in multiple 
ways~\cite{ZhangLeggett, Werner, BraatenPlatter} as well as verified experimentally~\cite{JET, JILA, KuhnleUniversal}. 

Recently, $C$ has also been found to determine the prefactor of the high-frequency power-law decay of 
correlators~\cite{SonThompson,TaylorRanderia}, as well as the right-hand sides of the shear- and bulk viscosity sum rules~\cite{TaylorRanderia}. 
The contact is therefore a central piece of information on the UFG in equilibrium as well as away from equilibrium, since it constrains 
several thermodynamic quantities with a single number. On the experimental side, $C$ has been shown to be central to radio-frequency
spectroscopy and laser photoassociation~\cite{Braaten}, as well as to govern the rate of decrease of low-energy atoms due to inelastic 
two-body scattering processes with a large energy release. The Tan relations (as well as the above-mentioned 
sum rules) remain valid at arbitrary $k_F^{}a$ as long as $k_F^{} r_0^{} \ll 1$. For further details and a comprehensive review,
see Ref.~\cite{Braaten}.

The calculation of $C$ itself, however, remains a challenge, as it depends on the intricate many-body dynamics of the unitary regime. 
In principle, $C$ can be extracted from any one of the Tan relations (as recently done in experiments~\cite{JILA}). 
%and thus the optimal 
%choice is dictated by the ability to evaluate them numerically. 
One of the simplest relations concerns the asymptotics of the momentum distribution, and asserts that
\beq
\label{Cdef}
C \equiv \lim_{k \to \infty} k^4_{} n_\sigma^{}(k), \quad
n_\sigma^{}(k) \equiv \langle \hat a^{\dagger}_{\sigma,k} \hat a^{}_{\sigma,k} \rangle,
\eeq
where $n_\sigma^{}(k)$ is the momentum distribution expressed as a thermal average, and the $\hat a^{\dagger}_{\sigma,k}$ and 
$\hat a^{}_{\sigma,k}$ denote creation and annihilation operators for particles of momentum~$k$ and spin~$\sigma$. 
If $n_\sigma^{}(k)$ is normalized to the particle number $N_\sigma^{}$, then $C$ is an extensive quantity with dimensions of momentum. 
We shall consider $C$ in units of $k^{}_F$ divided by the total particle number \mbox{$N = N^{}_\uparrow + N^{}_\downarrow$}. 

%Further details on the Tan relations, along with a comprehensive review, can be found in Ref.~\cite{Braaten}.

%as well as a comprehensive summary of the theoretical and experimental situation can be found in Ref.~\cite{Braaten}.

%%%%%%%%%%%%%%%%%%%

In this work, we focus on the momentum distribution of the homogeneous UFG and the extraction of $C$ via Eq.~(\ref{Cdef}), using 
a Quantum Monte Carlo~(QMC) approach which accounts fully for quantum and thermal fluctuations. On a 
spatial lattice, the Hamiltonian that captures the physics of the unitary limit can be written as 
\beq
\hat H \equiv 
\sum_{k} 
\frac{\hbar^2 k^2}{2m} 
\left ( 
\hat a^\dagger_{\uparrow k}
\hat a_{\uparrow k}^{}  + 
\hat a^\dagger_{\downarrow k} 
\hat a_{\downarrow k}^{} 
\right )
-\,g \sum_i \hat n_{\uparrow i}^{} \, \hat n_{\downarrow i}^{},
\eeq
%
%$\hat H \equiv \hat T + \hat V$, where the kinetic 
%energy operator $\hat T$ (in momentum space) is
%%
%\beq
%\hat T \equiv \hat T_\uparrow^{} + \hat T_\downarrow^{} =
%\sum_{k} 
%\frac{\hbar^2 k^2}{2m} 
%\left ( 
%\hat a^\dagger_{\uparrow k}
%\hat a_{\uparrow k}^{}  + 
%\hat a^\dagger_{\downarrow k} 
%\hat a_{\downarrow k}^{} 
%\right ),
%\eeq
%%
%and the potential energy operator $\hat V$ is given by the zero-range interaction
%%
%\beq
%\hat V \equiv -\,g \sum_i \hat n_{\uparrow i}^{} \, \hat n_{\downarrow i}^{},
%\eeq
%%
where $m$ is the mass of the fermions (henceforth set to unity), $g$ is the bare coupling, 
and $\hat n_{\sigma i}^{}$ denotes the number density operator for spin projection $\sigma$ at lattice position $i$. The 
equilibrium thermodynamical properties are obtained from the grand canonical partition function
\beq\label{Z_original}
\mathcal Z \equiv \Tr\,\exp[-\beta(\hat H \!-\! \mu \hat N)],
\eeq
where $\beta \equiv 1/k_B^{} T$, $\mu$ is the chemical potential, and 
\beq
\hat N \equiv \hat N_\uparrow^{} + \hat N_\downarrow^{} = \sum_i \hat n_{\uparrow i}^{} + \sum_i \hat n_{\downarrow i}^{}
\eeq
denotes the particle number operator.

In our QMC treatment, the system is placed on a $(3\!+\!1)$-dimensional Euclidean space-time lattice via a Suzuki-Trotter
decomposition of the Boltzmann weight in Eq.~(\ref{Z_original}), and the
interaction is represented via a Hubbard-Stratonovich (HS) transformation~\cite{HST}. As we focus on the spin-symmetric 
case, the fermion sign problem is absent. The resulting path 
integral formulation is an exact representation of the many-body problem of Eq.~(\ref{Z_original}), up to finite volume
and discretization effects. These may be addressed by varying the spatial lattice volume $V = N_x^3$ and the density $n$, such that
the thermodynamic and continuum limits are recovered as $V \to \infty$ and $n \to 0$. The latter requires great care, as too low densities 
imply a departure from the thermodynamic limit. We find that $n \simeq
0.03-0.05$ particles per unit volume yield results accurate to $\simeq 7\%$ at finite temperature, and to $\leq 5\%$ at $T=0$.

\begin{figure}[t]
\epsfig{file=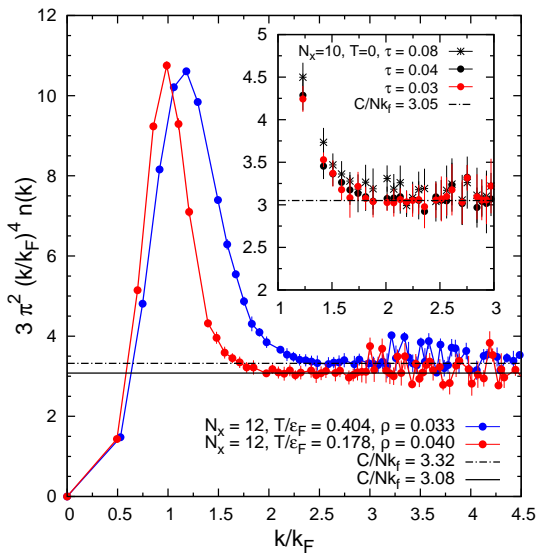, width=0.84\columnwidth}
\caption{(Color online) Plot of $3 \pi^2 (k/k_F^{})^4_{} n(k)$ for $N_x^{} = 12$ as a function of $k/k_F^{}$ at
$T/\epsilon_F^{} = 0.178$ and $0.404$. The ``plateaux'' at large $k/k_F^{}$ give 
the (intensive) dimensionless quantity $C/(N k_F^{})$. At low $T/\epsilon_F^{}$, the asymptotic region is 
reached at $k/k_F^{} \simeq 2$. Inset: $N_x^{} = 10$ results at $T = 0$ showing only slight dependence on the Suzuki-Trotter
step $\tau$.
\label{Fig:k4nk}}
\end{figure}

Our lattice formulation is very similar to Ref.~\cite{BDM}, but differs in at least three notable 
aspects. Firstly, we determine the bare lattice coupling constant $g$ corresponding to the unitary regime by using L\"uscher's 
formula~\cite{Luescher} as in Ref.~\cite{LeeSchaefer}. This procedure yields $g \simeq 5.14$ in the unitary limit. 
Secondly, we use the compact, continuous HS transformation
\bea
\label{LeeHS}
\exp\left(\tau g \,\hat n_{\uparrow i}^{} \hat n_{\downarrow i}^{}\right) &=& 
\frac{1}{2\pi} \int_{-\pi}^{\pi} d\sigma_i^{}
\left[1 + B \sin(\sigma_i^{}) \, \hat n_{\uparrow i}^{}\right] \nonumber \\
&& \times \left[1 + B \sin(\sigma_i^{}) \, \hat n_{\downarrow i}^{}\right],
\eea 
where $\sigma_i^{}$ (not to be confused with the spin projection) is the HS auxiliary field, with $B^2/2 \equiv \exp(\tau g) - 1$, 
and $\tau$ denotes the lattice spacing in the imaginary time direction. We find that $\tau \simeq 0.05$ is sufficiently small to render 
discretization errors from the Suzuki-Trotter decomposition insignificant (see also Fig.~\ref{Fig:k4nk}). The above representation 
(referred to as ``Type~4'' in Ref.~\cite{DeanLee}) was found to be superior with respect to acceptance rate, decorrelation
and signal-to-noise properties than the more conventional unbounded and discrete forms~\cite{Hirsch}. Thirdly, we update the auxiliary 
field $\sigma$ by using the Hybrid Monte Carlo (HMC) algorithm~\cite{DuaneGottlieb} (familiar from Lattice 
QCD), which combines the Metropolis algorithm with deterministic Molecular Dynamics. 
%The HMC algorithm is exact even at finite MD timestep, provided that the evolution is symplectic and reversible. 
Our implementation of the HMC algorithm enables global updates at all temperatures and lattice sizes, and scales approximately as $\sim V^2_{}$ as a 
function of the spatial lattice volume, to be contrasted with the $\sim V^3_{}$ scaling of approaches based on local updates. 
%Thus, HMC allows us to efficiently sample to configuration space of the HS auxiliary field. 
%Additionally, it should be noted that localized moves become increasingly ineffective, for both large lattices and 
%low temperatures. This problem is largely avoided in our implementation of HMC, in particular because we do not resort to iterative Conjugate 
%Gradient~(CG) inversion during the MD evolution. 
%The appropriate choice of HS transformation is central to the HMC approach, as the MD evolution 
%is sensitive to the appearance of ``exceptional configurations'' which may lead to numerical instabilities.

%%%%%%%%%%%%%%%%%%%

%
\begin{figure}[t]
\epsfig{file=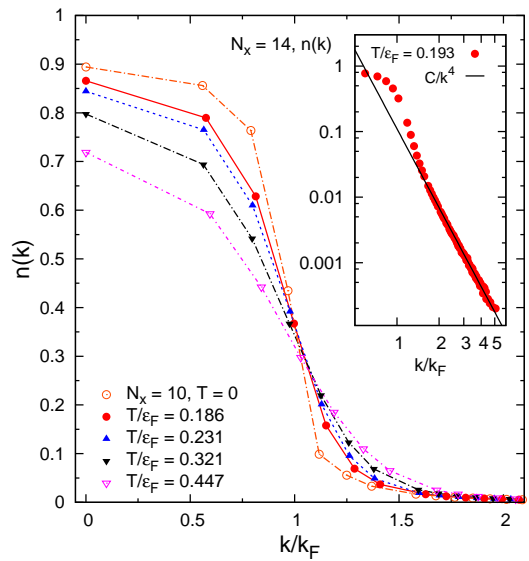, width=0.85\columnwidth}
\caption{(Color online) Momentum distribution $n(k)$ from QMC for $N_x^{} = 10$
as a function of $k/k_F^{}$, for various temperatures ranging 
from zero to $T/\epsilon_F^{} \simeq 0.5$. The curves are intended as a guide to the eye, and the statistical errors are
the size of the symbols. Inset: $n(k)$ for $N_x^{} = 14$ in a log-log scale, showing the asymptotic $\sim k^{-4}_{}$ behavior.
\label{Fig:nk}}
\end{figure}

\begin{figure}[t] 
\epsfig{file=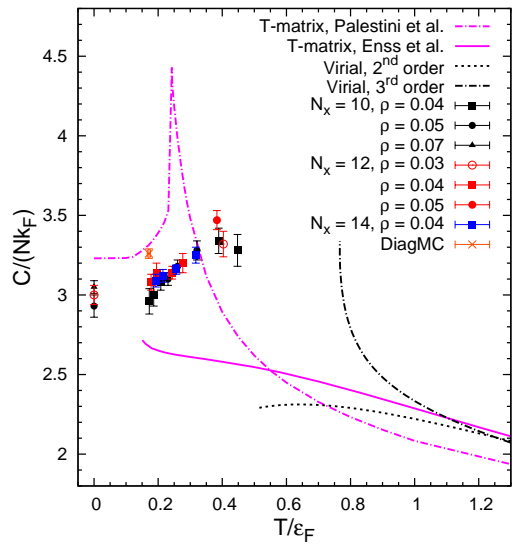, width=.84\columnwidth}
\caption{(Color online) Summary of QMC results for $C/(N k_F^{})$ as a function of $T/\epsilon_F^{}$, as determined 
from the large $k/k_F^{}$ behavior of $n(k)$. The errorbars are dominated by systematics related to the
residual fluctuations in the plateaux, as exhibited in Fig.~\ref{Fig:k4nk}. 
Also shown are the t-matrix calculations of Ref.~\cite{Palestini, Enss}, the virial expansion of Ref.~\cite{Hu}
and the diagrammatic Monte Carlo result of Ref.~\cite{Goulko}.
%Also shown are the $t$-matrix calculations of 
%Refs.~\cite{Palestini, Enss} and the virial expansion of Ref.~\cite{Hu}.
\label{Fig:Contact}}
\end{figure}

We have performed calculations at $T = 0$ as well as $T/\epsilon_F^{} > 0$, in the former case using an approach 
similar to Ref.~\cite{DeanLee}. Our main results correspond to $40-50$ particles at $N_x^{} = 10$ and 
$70-80$ particles at $N_x^{} = 12$, in addition to limited data for $N_x^{} = 14$.
In Fig.~\ref{Fig:nk}, we show the momentum distribution $n(k)$
as a function of $T/\epsilon_F^{}$. We have computed $n(k)$ by averaging over the angular directions on the lattice as well as over 
the imaginary-time slices. In this way, we find that $\sim 200$ uncorrelated auxiliary field samples for each 
datapoint gives excellent statistics for $n(k)$. Multiplying $n(k)$ by $k^4_{}$, as plotted in Fig.~\ref{Fig:k4nk}, we find a peak 
at $k\simeq k^{}_F$ and a leveling out at high momenta, with the asymptotic regime setting in at $k \simeq 2 k_F^{}$ at the lowest
temperatures. It is fortuitious that the asymptotic regime sets in at such low momenta, as there is no obvious reason for this to be the case. 
It is then possible to study the temperature dependence of this ``plateau'', which allows 
for a determination of the contact $C/(Nk_F^{})$ as a function of $T/\epsilon_F^{}$.
These results are given in Fig.~\ref{Fig:Contact}, together with a comparison with available theoretical analyses. Our results
indicate that $n(k)$ follows the expected $\sim k^{-4}_{}$ dependence very accurately up
to at least $k \simeq 4 k_F^{}$, at which point the signal deteriorates due to lattice artifacts.

%Indeed, if the particle number is much reduced at constant volume, the momentum distribution eventually approaches that of an empty system, 
%in which $C/(Nk_F^{})$ obviously vanishes.

%The maximum in $C/(Nk_F^{})$ at $T/\epsilon_F^{} \sim 0.3$ indicates enhanced short range correlations. While these have been related 
%to the existence of a pseudogap, and therefore to the fact that Fermi gases close to unitarity remain strongly correlated even above the critical 
%temperature $T^{}_c \simeq 0.15\epsilon_F^{}$, the contact bears little connection to the pairing correlations that lead to superfluidity. 
%As mentioned above the physical meaning of the contact 

The value of $C$ in the ground state can be computed via Diffusion Monte Carlo (DMC) calculations, as first done in 
Ref.~\cite{Lobo} using density-density correlations, which yielded \mbox{$C(T=0)/(Nk_F^{}) \simeq 3.4$}, up to errors associated with
fixing the nodes of the wavefunction. A more recent and comprehensive DMC calculation~\cite{Stefano} came to the same conclusion 
using the equation of state, the momentum distribution and the density-density correlation. In contrast, our present results indicate 
that \mbox{$C(T=0)/(Nk_F^{}) \simeq 2.95 \pm 0.10$}. The cause of this disagreement is being explored.
The main sources of uncertainty in our determination of $C/(Nk_F^{})$ are due to finite density effects. While we find that such effects tend to
overestimate $C/(Nk_F^{})$ as well as degrade the formation of an asymptotic $\sim k^{-4}_{}$ tail in $n(k)$ at larger 
values of $T/\epsilon_F^{}$, larger lattices are needed in order to maintain the thermodynamic limit at lower
densities.

%However, the need for lower densities should
%be balanced against the requirement of larger lattice volumes, in order to maintain the thermodynamic limit. 

%

The temperature dependence of $C$ at unitarity was first determined analytically in Ref.~\cite{Yu}, who considered two different limits. 
At very low temperatures $T \ll T_c^{} \simeq 0.15 \,\epsilon_F^{}$, the dominant excitations are of phononic origin, and the 
$T$-dependence of $C$ is of the form $C/(Nk_F^{})\! \propto \! \left({T}/{\epsilon_F}\right)^4_{}$. On the other hand, at very high 
temperatures $T \! \gg \! \epsilon_F^{}$, one finds $C/(Nk_F^{})\! \simeq \! 16/3\,(\epsilon_F^{}/T)$ within the second-order virial expansion. 
An interpolation between these limits then suggests that $C(T/\epsilon_F^{})$ should present a maximum for $T\! \sim \! \epsilon^{}_F$. 
Recently, $C$ has also been computed using two different types of $t$-matrix 
approximations~\cite{Palestini, Enss}, as well as a third-order virial expansion~\cite{Hu}. The latter 
has shown evidence for convergence of the virial expansion down to $T \sim \epsilon^{}_F$. 
In light of these findings and upon analysis of various model calculations at low~$T$, 
Ref.~\cite{Hu} conjectured that the contact is likely a monotonically decreasing function 
of $T$, except possibly in the phononic regime at very low~$T$. While the virial expansion is on solid ground at high~$T$, where it agrees 
with the $t$-matrix approaches of Refs.~\cite{Palestini, Enss}, the actual $T$-dependence in the strongly correlated \mbox{low-$T$} regime 
has remained an open question, particularly since the UFG is strongly correlated even above 
$T_c^{}$~\cite{Pseudogap}.

Our results show that $C$ grows with $T$ well beyond the superfluid phase, and are suggestive of a maximum
$C_\mathrm{max}^{} \simeq 3.4$ at $T/\epsilon_F^{} \simeq 0.4$. This scenario is in qualitative agreement with 
Ref.~\cite{Yu}, as well as the $t$-matrix calculation of Ref.~\cite{Palestini}. As $C$ measures the number of particle pairs (of both
spins) whose separation is small, the appearance of a maximum indicates an enhancement in such short-range correlations. 
This may be a result of local pairing order~\cite{Palestini}, which in turn suggests that $C_\mathrm{max}^{}$ is directly 
related to pairing above $T^{}_c$, {\it i.e.} to a pseudogap. We find the scale at which the $k^{-4}_{}$ law sets 
in (see Fig.~\ref{Fig:k4nk}) to be $k \simeq 2 k^{}_F$ at finite $T/\epsilon_F^{}$ and somewhat lower for the ground state, in agreement 
with Ref.~\cite{JILA}. This universal property of the unitary limit characterizes the ``healing distance'' of the 
two-particle boundary condition on the many-body wavefunction, and therefore 
separates the microscopic properties from the universal macroscopic aspects of the unitary regime.
Direct comparison of our data with ultracold atom experiments can be achieved by means of the virial expansion and the Local 
Density Approximation (LDA). While we defer this issue to a follow-up paper, we note that in light of 
the work of Ref.~\cite{Kuhnle}, the features of $C(T/\epsilon^{}_F)$ found in this study are unlikely to conflict with
current experiments.

In summary, we have computed the momentum distribution $n(k)$ and the contact $C/(Nk_F^{})$ for the UFG at zero and 
finite $T/\epsilon_F^{}$, using the auxiliary field QMC method in conjunction with the HMC
algorithm. While the ground-state momentum distribution was first determined via DMC calculations in Ref.~\cite{Astrakharchik}, our results 
represent the first fully non-perturbative calculation of $n(k)$ free of uncontrolled approximations.
We find that the contact at $T = 0$ assumes the value $\simeq 2.95 \pm 0.10$ and increases as a function of $T/\epsilon_F^{}$ in the low- 
and intermediate temperature regimes that we have explored, which is consistent with the phononic scenario. Notably, DMC
calculations find a somewhat larger value of $C/(Nk_F^{}) \simeq 3.4$, while the analytic 
approach of Ref.~\cite{Haussmann}, which interpolates smoothly between the strong- and weak-coupling limits, yields $C/(Nk_F^{}) \simeq 3.0$ 
which is consistent with our data. Our results complement the calculations of Refs.~\cite{Yu,Hu,Palestini,Enss}, 
and are suggestive of a maximum in $C/(Nk_F^{})$ at $T/\epsilon_F^{} \simeq 0.4$, which agrees qualitatively with 
Ref.~\cite{Palestini} but disagrees with Ref.~\cite{Enss}. While calculations at higher $T/\epsilon_F^{} \sim 1$ are feasible, an improved
understanding of the finite density effects is clearly called for.

%we have 
%found that the finite density effects become potentially very large at higher temperatures, such that larger lattices and lower densities 
%are called for.

%%%%%%%%%%%%%%%%%%%

\begin{acknowledgments}

We thank R.~J.~Furnstahl for encouragement and A.~Bulgac, J.~Carlson, S.~Gandolfi, A.~Gezerlis and K.~Schmidt for 
instructive discussions and comments. We are also grateful to T.~Enss, H.~Hu, and F.~Palestini for giving us access to their 
respective results. We acknowledge support under U.S. DOE Grants 
No.~DE-FG02-00ER41132 and DE-AC02-05CH11231, UNEDF SciDAC 
Collaboration Grant No.~DE-FC02-09ER41586 and NSF Grant No.~PHY--0653312. 
This study was supported, in part, by the Academy of Finland through its Centers of Excellence Program (2006 - 2011), 
the Vilho, Yrj\"o, and Kalle V\"ais\"al\"a Foundation of the Finnish Academy of Science and Letters, and the Waldemar von Frenckell 
and Magnus Ehrnrooth Foundations of the Finnish Society of Sciences and Letters. Part of this work was performed 
using an allocation of computing time from the Ohio Supercomputer Center.

\end{acknowledgments}

%%%%%%%%%%%%%%%%%%%

%%%%%%%%%%%%%%%%%%%

\end{document}